\documentclass[useAMS,usenatbib]{mn2e}

\usepackage{txfonts,graphicx,amssymb}
\usepackage[normalem]{ulem}

\newcommand\aap{A{\&}A}

\newcommand\apj{ApJ}

\newcommand\mnras{MNRAS}

\def\pasp{PASP}
\def\procspie{Proc.~SPIE}

\def\vsini{\upsilon\sin i}

\title[Dynamos in M dwarfs]
{Towards understanding dynamo action in M dwarfs}

\author[D.~Shulyak, D.~Sokoloff, L.~Kitchatinov, D.~Moss]
       {D.~Shulyak$^1$, D.~Sokoloff$^2$, L.~Kitchatinov$^{3,4}$, D.~Moss$^5$ \\
$^1$Institute of Astrophysics, Georg-August University, Friedrich-Hund-Platz 1, D-37077 G\"ottingen, Germany\\
$^2$Department of Physics, Moscow University, 119899, Moscow, Russia  \\
$^3$Institute of Solar-Terrestrial Physics, PO Box 291, Irkutsk 664033, Russia\\
$^4$Pulkovo Astronomical Observatory, St. Petersburg, 196140, Russia\\
$^5$School of Mathematics, University of Manchester, Oxford Road,
Manchester, M13 9PL, UK
}

\begin{document}

\date{Accepted  .... Received  ....; in original form }

\pagerange{\pageref{firstpage}--\pageref{lastpage}}
\pubyear{2014}

\maketitle

\label{firstpage}

\begin{abstract}
Recent progress in observational studies of magnetic activity in M dwarfs urgently requires support from ideas
of stellar dynamo theory. We propose a strategy to connect observational and theoretical studies.
In particular, we suggest four magnetic configurations that appear relevant to dwarfs from the viewpoint of the
most conservative version of dynamo theory, and discuss observational tests to identify the configurations
observationally. As expected, any such identification contains substantial uncertainties. However the
situation in general looks less pessimistic than might be expected. Several identifications between the
phenomenology of individual stars and dynamo models are suggested. Remarkably, all models discussed predict
substantial surface  magnetic activity at rather high stellar latitudes. This prediction looks unexpected from the viewpoint
of our experience observing the Sun (which of course differs in some fundamental
ways from these late-type dwarfs).
We stress that a fuller understanding of the topic requires a long-term (at least 15 years)
monitoring of M dwarfs by Zeeman-Doppler imaging.
\end{abstract}
\section{Introduction}
\label{intro}
The intensive investigation of stellar magnetic activity has already a history of
about 40 years, and many important results
have been obtained \citep{HN87,B09,S11,D11,Vea14,KO14}.
On one hand, this activity is
believed to be similar to some extent to the solar cyclic activity. In particular, the magnetic fields of many types of
stars are believed to be excited by stellar dynamo action, driven by
the joint action of differential rotation and an
additional mirror-asymmetric driver which produces poloidal magnetic field from toroidal. The physical nature of this
mechanism is still debated for the Sun, and there are even more uncertainties
for other stars. However we have no intention to enter this discussion
now and for the sake of simplicity refer to this factor as the $\alpha$-effect,
whatever its physical nature.

On the other hand, magnetic activity in any particular star can look quite
distinctively different from the solar
case:  starspots on the most investigated magnetically active stars are much larger than on the Sun, covering
a larger part of the surface than sunspots, etc.
These specific features should somehow be related to the properties
of stellar dynamos. However in practice any such relation is ill-understood.

In particular, the relation between the magnetic phenomenology of M dwarfs and
the underlying stellar dynamo action is a topic of
continuing discussions at scientific meetings. Experts in observational investigations of  starspots on
M dwarfs have stressed many times that some input from dynamo theory to
the problem is very desirable. There is a general feeling that there
is a basic difference between dynamo action in M dwarfs
which are fully convective stars, and in the Sun which has a relatively
thin convective shell. As far as we can recall, it was
E.~Ergma who attracted the attention of one of us (Dmitry Sokoloff)
to the problem as early as 1993.

There are many theoretical models that explore
various details of stellar
dynamo action for a range of stellar types, including M dwarfs \citep[e.g.,][]{CK06,B08}.
3D dynamo models give very important information concerning magnetic activity of fully convective stars. In
particular, the models clarify spot formation mechanism \citep{2015A&A...573A..68Y} and
predict the location of dipole-dominated M-dwarfs \citep{schrin12,gast12,2013A&A...549L...5G,schrin14}
in terms of their rotation velocities and convection efficiency, 
giving in this way an observational test for verification of the model.
However, sometimes the predicted magnetic configurations
depend significantly on various small and quite uncertain factors,
such as the boundary conditions at the stellar surface;  it is thus
very difficult to adequately reproduce all these details in a particular dynamo model.

Therefore, although considerable progress has been made in
developing comprehensive 3D MHD simulations of fully or nearly fully
convective stars \citep[e.g.][]{dSB06,B08,schrin12,gast12,schrin14,2015A&A...573A..68Y},
these models are not yet fully satisfactory.
For instance, applied to the Sun the 3D models still have difficulties
in tuning parameters to reproduce the observed butterfly diagram.

Depending on assumptions and approximations used in numerical simulations, 
modern dynamo models may predict different dymano states.
For instance, for fully convective stars it is often suggested that dynamos will be of
$\alpha^2$ type, and often non-axisymmetry is predicted.
However, the steady $\alpha^2$ solutions seem to be not fully 
in accordance with observations, where axisymmetric and
non-axisymmetric configurations are found in very similar
objects \citep[see Sect.~\ref{observ}, also][]{gast14}.
Note that relatively weak toroidal field is often taken as an indicator of
non-axisymmetry. On the other hand, as was shown in e.g., \cite{schrin12},
oscillating $\alpha^2$-dynamos are also possible under certain conditions.

Furthermore, very recent dynamo models predict that dynamo can be
bistable, the eventual stable state depending on initial conditions.
This could explain the observed
distributions of the magnetic field geometries among objects with similar
parameters \citep[see e.g., ][]{schrin14}.

One of the difficulty in using 3D models for the extensive exploration
of stellar dynamos is the requirenment of large computational resources, 
making it difficult to use outside of a substantial
programme, and limiting possible exploration of parameter space.
As alternative, the ``hybrid'' models that use transport coefficients deduced from MHD simulations
in mean field models may provide a more satisfactory way
forward \citep[e.g.,][]{dube13,pip13}. Nevertheless, in spite of a wide variety of
available models, none of them provides a complete understanding of dynamo action in low mass stars.

The above considerations suggest that progress in theoretical modelling
and observations has led to a rather strange situation where
quite amazing advances in the last 40 years have brought
rather less physical understanding of the problem than
might have been anticipated. Our feeling is that coordination of efforts
between observers and dynamo modellers is urgently needed
to further progress in the topic.

We envisage the format of the coordination to be as follows.
A risk has to be taken and a particular approach to dynamo modelling
has to be accepted as a starting point.
A list of magnetic configurations that can be expected for M dwarfs in the
framework of this basic model has to be constructed,
together with observational tests capable of distinguishing the configurations
observationally. If possible, the existing observational data have
to be used to identify the configurations, and the
perspective of future observations should be discussed. The point is that the full inspection of the list may (and
will) be quite demanding on telescope time and other material resources,
and the community will have to make a
reasonable decision on how to manage resources in order to
achieve this goal. Further steps in  development of the model would have
to be motivated by particular difficulties in interpretation of observations.

The aim of this work is to explore the magnetic configurations that
can be excited by a more or less standard mean-field dynamo model based on 
differential rotation ($\Omega$-effect) and mirror-asymmetric convective
motions (the $\alpha$-effect), and then to
follow the above plan as far as it is possible at the moment.
We appreciate that several important points
of our (as any other) dynamo model
are debatable and can only, at best, partially represent the
true physics. In particular, we recognize that in other models the differential
rotation may be substantially reduced compared to the model we consider
in this work. We stress however, that this paper is more
a demonstration of principle than an attempt to offer a definitive
resolution of any particular issue, and should be viewed as such.
We do not attempt to give a complete review of the field, but rather just
to indicate some key points and references.
Of course, the parameters and details of the models are likely to
be modified in course of progress in understanding of the topic.

This paper is in some sense related to the recent paper by \cite{ketal14} who, however, approached the
topic from a quite different viewpoint. They suggested an explanation for  the observed
phenomenology of magnetic activity of M dwarfs in a way analogous to the solar case: that
explanation does not require the idea of dynamo bistability. Here,
however, we do not exclude the latter in principle and address
the problem from an alternative viewpoint, and discuss the richness
of dynamo models for M dwarfs that have more or less comparable
complexity. As we will show below, four rather than two configurations look possible and we suggest how to clarify
which (if any) are really of potential interest, and whether something more complex is needed.

\section{The dynamo model}

Dynamo theory explains stellar magnetic activity by two basic effects:
i) poloidal field production from toroidal field by cyclonic motions
(the $\alpha$-effect) and ii) conversion of poloidal field back to toroidal by
the action of differential rotation (the $\Omega$-effect) and possibly
by the $\alpha$-effect as well. The efficiency of the two effects in
generating magnetic fields are measured by the dimensionless parameters
\begin{equation}
    C_\Omega = \frac{\Delta\Omega R_\mathrm{star}^2}{\eta_{_\mathrm{T}}}
    \label{C_omega}
\end{equation}
for the $\Omega$-effect ($\Delta\Omega$ is a measure of the angular velocity variation within the convection zone
and $\eta_{_\mathrm{T}}$ is the eddy magnetic diffusivity), and
\begin{equation}
    C_\alpha = \frac{\alpha_0 R_\mathrm{star}}{\eta_{_\mathrm{T}}},
    \label{C_alpha}
\end{equation}
for the $\alpha$-effect \citep{kr80}, where $\alpha_0$ is a representative value.

For our illustrative purposes we use the same dynamo model as
\citet{ketal14} and therefore describe it only briefly.
The model is of $\alpha^2\Omega$-type, i.e. taking the toroidal field generation
by the $\alpha$-effect into account. The differential rotation in M dwarfs is small \citep{betal05}
and it is often neglected in models of their dynamos, thus giving $\alpha^2$-dynamos.
We note that it is possible to obtain similar magnetic field configurations
with or without differential rotation. 
For instance, axisymmetric fields have been obtained in recent simulations with pure 
$\alpha^2$-dynamos \citep[e.g.,][]{schrin12}. 
On the other hand, mean-field models of $\alpha^2$-dynamos
by \citet{CK06} (which allow for the rotationally induced anisotropy 
of the $\alpha$-effect) predict non-axisymmetric global fields similar to an equatorial dipole.
But if differential rotation is present, it may oppose the non-axisymmetry by converting 
azimuthal variations into small-scale radial or meridional structure, enhancing diffusive decay \citep{MB95}.
Because the results of simulations are very sensitive to approximations and assumptions made,
in the mean-field models that we use in this paper we explicitly
include the contribution from differential rotation.
The point here is that although $\Delta\Omega$ decreases with decreasing stellar mass,
the eddy magnetic diffusion $\eta_{_\mathrm{T}}$ decreases even faster so that the parameter
$C_\Omega$ (Eq.~\ref{C_omega}) {\em increases}. This trend was found in mean-field models of
differential rotation by \citet{ko11}.
Note again that this model does not prescribe the eddy transport
coefficients but estimates them in terms of the entropy gradient, the entropy being
a dependent variable of the model.
In particular, the value $\nu_{_\mathrm{T}} \simeq 1.2\times 10^{11}$\,cm$^2$s$^{-1}$
was estimated for the middle radius $r = R_\mathrm{star}/2$ of a $0.3M_\odot$ star
rotating with a period of 10 days ($\nu_{_\mathrm{T}}$ varies moderately with depth).
We shall use the differential rotation computed for this fully convective star in our dynamo model.
The eddy magnetic diffusion is of the same order of magnitude as the eddy viscosity \citep{YBR03},
so that the magnetic Prandtl number,
\begin{equation}
    P_\mathrm{m} = \nu_{_\mathrm{T}}/\eta_{_\mathrm{T}},
    \label{P_m}
\end{equation}
does not differ much from unity. Taking $P_\mathrm{m} = 1$ gives
$C_\Omega \simeq 290$ which is not small in spite of a rather small
differential rotation of about 1.5\% \citep[see Fig.1 in][]{ketal14}.

The value of $C_\alpha$ (Eqn.~\ref{C_alpha}) is more difficult to estimate.
The origin of the $\alpha$-effect is not certain for the Sun \citep{C14} 
and even less so for M-dwarfs. Order of magnitude estimation is nevertheless possible. 
The Rossby number $\mathrm{Ro} = P_\mathrm{rot}/\tau$ is small for the $0.3M_\odot$ star 
considered with $P_\mathrm{rot} = 10$~days ($\tau$ is the convective turnover time with 
a characteristic value of a few months \citep[e.g.,][]{2010ApJ...721..675B}). 
The $\alpha_0$ parameter for this case of convection strongly affected by rotation is expected 
to be close to its maximum possible value of the convective velocity $u_\mathrm{c} \sim 3\eta /\ell$; 
$\ell$ being the mixing length. The estimate $C_\alpha \sim 3R_\mathrm{star}/\ell$ follows 
from Eq. (\ref{C_alpha}). With the plausible value of $R_\mathrm{star}/\ell \sim 10$, 
this gives $C_\alpha \sim 30$, which is not much above the critical $C_\alpha$ values 
for onset of dynamo action in our model (a stellar structure model gives $R_\mathrm{star}/\ell = 4.3$ 
for the middle radius $r = R_\mathrm{star}/2$ leading to a somewhat smaller $C_\alpha$).

It may be expected, however, that back reaction of generated magnetic fields on motion affects 
primarily the $\alpha$-effect \citep{BS05}. The majority of mean-field dynamo models employ 
the $\alpha$-quenching as the only nonlinearity. The quenching reduces $C_\alpha$ to close 
to the threshold value for onset of dynamo action.

Note that estimates of $C_{\alpha}$ and $C_{\Omega}$ suggest the importance of differential rotaion
only in the models that we have chosen to use in this paper. 
This, of course, does not exclude the pure $\alpha^2$-dynamo 
action in M dwarfs, as was stressed above. 
However, the theoretical ability to predict dynamo 
governing parameters does not allow one to obtain firm predictions 
and we believe that observations may play a decisive role here.

We apply our kinematic dynamo model to estimate the threshold values of $C_\alpha$ for magnetic fields of different equatorial and axial symmetries.
Our model assumes uniform diffusion and that the $\alpha$-effect is
uniform with radius and varies
as $\cos\theta$ with co-latitude $\theta$, $\alpha = \alpha_0\cos\theta$.
The model solves the eigenvalue problem for the mean-field dynamo equations \citep[cf., e.g.,][]{kr80}
numerically by applying a grid point method in radius and a latitudinal
expansion in Legendre polynomials. Computations were performed on a  uniform radial grid of 301 points, and with 40 Legendre polynomials in the
latitudinal expansion. Vacuum conditions were applied at the outer boundary.
These conditions demand that the toroidal field vanishes at the surface.
We cannot exclude however the possibility that the vacuum
boundary condition is not the whole story. It is possible to have a field above the surface which is force-free, with finite toroidal field, e.g.
\citet{m66}, later \citet{mw76}, see also the discussion in \citet{ms09}.

Our model takes $\alpha$ to be independent of radius.
Arguing from a possible correlation with the kinetic helicity (which is
often used as a proxy for the $\alpha$-effect), it might
be appropriate to take $\alpha$ as an increasing function of radius. As stressed
above, we only wanted to provided an illustrative model on which to hang our
arguments, rather than to provide an extensive exploration of parameters,
and so did not pursue this.

For numerical reasons the model requires an inner radial boundary.
This artificial boundary was imposed at $r_\mathrm{i} = 0.1 R_\mathrm{star}$ ($R_\mathrm{star}$ = 212 Mm).
Conditions for an interface with a perfect conductor were imposed at the inner boundary.

Our toy model has a distinctly non-cylindrical rotation profile \citep[Fig.~1 of][]{ketal14},
in contrast to predictions of 3D simulations of rapidly rotating
models with strong dynamo action \citep[e.g.][]{B08}. Feedback from Lorentz torques can reduce differential
rotation.
Weaker fields will merely modify the
underlying rotation field \citep[e.g.][]{MB2000}. 

We appreciate that a number of potentially significant changes 
connected with the inclusion of non-linear effects (alpha-quenching, Lorentz force, etc.) 
could be made to the model and this could affect the saturated states \citep[e.g.,][]{MB2000,B08}. 
To avoid these complications, below we consider only linear models.
As was stressed above, we prefer to start our exploratory work with simple models that will
be advanced once required by the analysis of observed data.

\section{Dynamo excited magnetic field configurations}
\label{mod}

We use the standard notations S$m$ and A$m$ for global modes of magnetic fields
\citep[cf.][]{kr80}, where the first letter S or A denotes
field configurations that are symmetric or antisymmetric,
respectively, with respect to the equator. In this notation,
$m$ is the azimuthal wave number. A0 and S0, therefore denote axisymmetric
global fields, which can be steady or oscillatory. A1 and S1 are
non-axisymmetric modes with $m=1$;  each field component changes sign twice over an entire longitudinal circle.
The modes with $m \geq 2$ normally have dynamo excitation thresholds that
are too high for us to consider them as a probable outcome of global dynamo
action in M dwarfs.
All the global modes combine poloidal and toroidal field components.
The meaning of toroidal and poloidal fields for axisymmetric modes is well
known. Lines of toroidal fields in non-axisymmetric modes lay on concentric spherical surfaces.
Poloidal non-axisymmetric fields have toroidal vector potentials and are supported by toroidal currents
\citep[see][ for more details]{C60}.

\begin{figure}
\begin{center}
\includegraphics[width=0.45\textwidth]{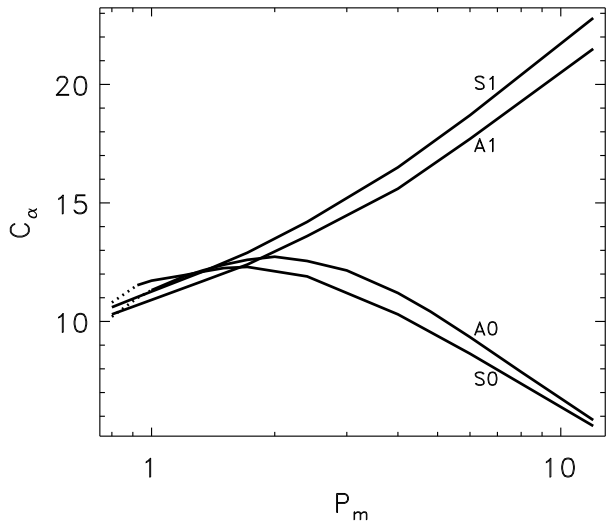}
\end{center}
\caption{\label{diff}
A dynamo model for a M dwarf:  marginal values of $C_\alpha$ (Eqn.~\ref{C_alpha})
for axisymmetric modes (A0) and non-axisymmetric $m=1$ modes (A1) of dipolar
symmetry, and axisymmetric modes (S0) and
non-axisymmetric modes (S1) of quadrupolar symmetry.
Dots indicate steady axisymmetric modes and the solid line shows the oscillatory modes.}
\end{figure}

Thinking straightforwardly, we have to conclude that the modes that can
be excited for the lowest $C_\alpha$, i.e.
the A1 mode for $P_\mathrm{m} < 1.1$ and S0 for  $P_\mathrm{m} >1.1$, are the modes likely to be excited by dynamo action in M- dwarfs.
The point however is that the difference in excitation condition between the modes S1 and A1 for lower
$P_\mathrm{m}$ and between the modes A0 and S0 for larger  $P_\mathrm{m}$
respectively is quite small and it would
be unrealistic to think that contemporary dynamo models are sufficiently near physical reality
to support such distinctions.
Thus we allow that any one of these modes might be excited in a given M dwarf. We assume that the probabilities of getting
any of the configurations for a given
$P_\mathrm{m}$ are comparable for the members of the pair.

To be definite, we ignore at least for the time being the non-oscillating modes shown by dots in
Fig.~\ref{diff}. We stress that the oscillating nature of the nonaxisymmetric
modes S1 and A1 is non-physical:
there is a rotating frame in which each of these magnetic configuration is non-oscillating. In other words,
oscillation of
the modes S1 and A1 is just a result of  longitudinal drift of the field patterns, viewed from a given line of sight.

\begin{figure}
\begin{center}
\includegraphics[width=0.45\textwidth]{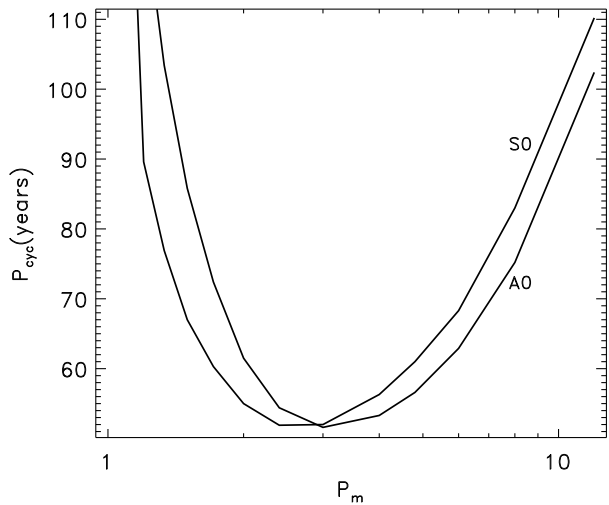}
\end{center}
\caption{\label{dwarfdyn}
The period of activity cycles for truly oscillating modes (to be compared with the 22-year solar activity cycle).}
\end{figure}

\begin{figure}
\begin{center}
\includegraphics[width=0.45\textwidth]{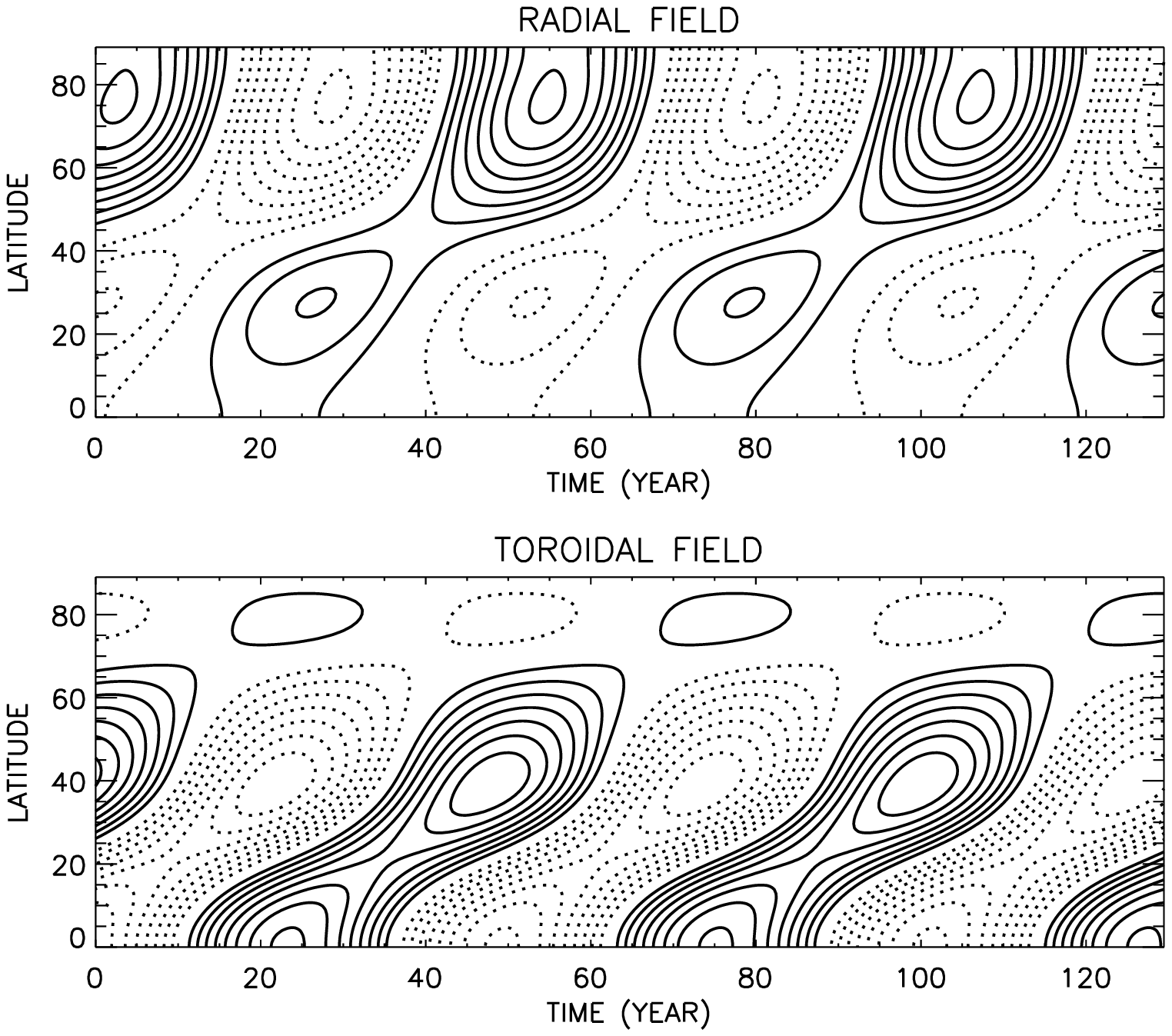}\\
\includegraphics[width=0.45\textwidth]{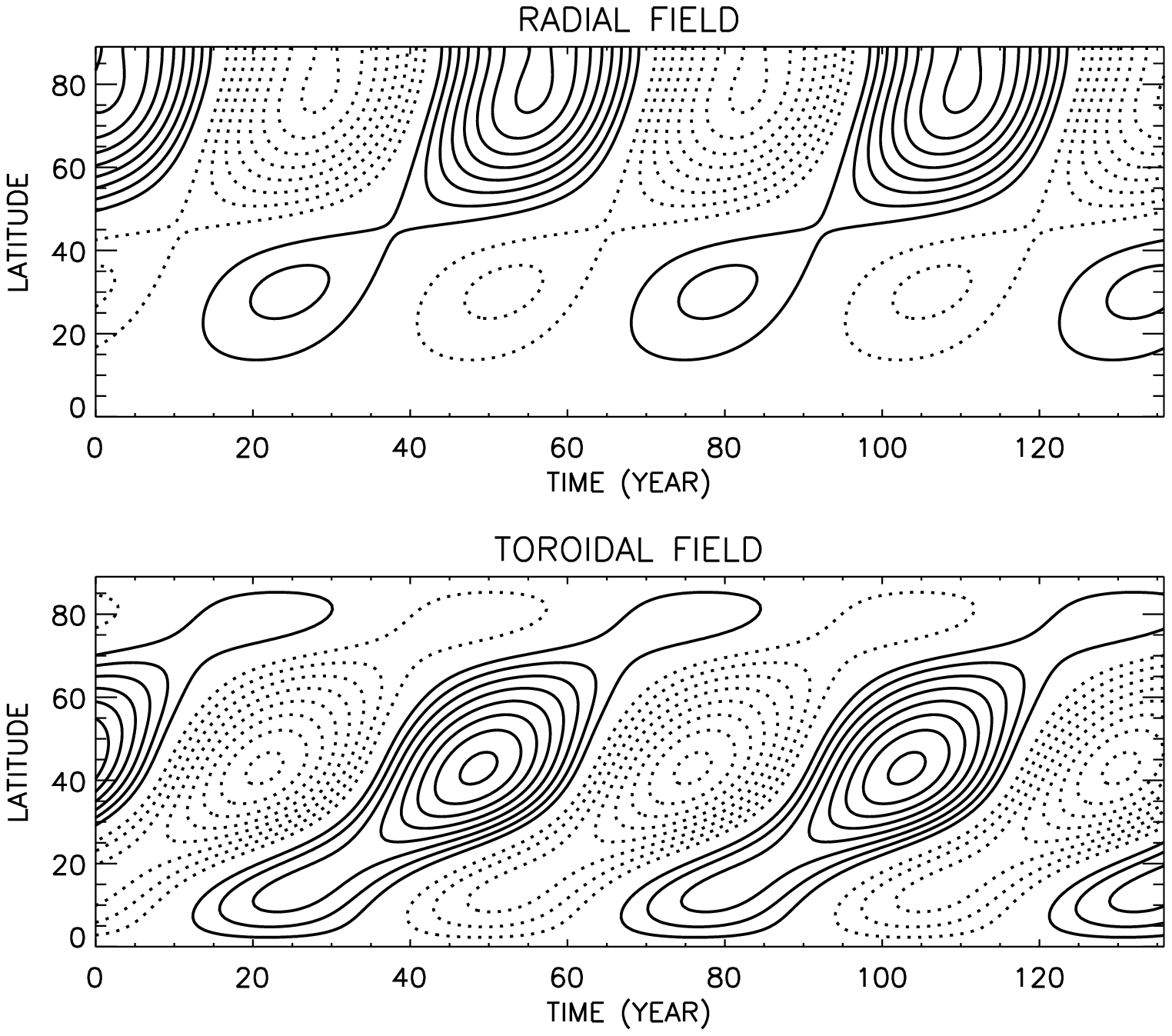}
\end{center}
\caption{\label{buttdiag}
Butterfly diagrams for truly oscillating modes: upper pair - mode S0,
lower pair - mode A0. Patterns of radial field at the surface
and toroidal field at small depth are shown. Pm = 2.4.}
\end{figure}

The period of activity cycles for truly oscillating modes are shown in Fig.~\ref{dwarfdyn}. Butterfly diagrams for these
modes are shown in Fig.~\ref{buttdiag}. Note that the activity waves propagate polewards.
\footnote{Application of the same model to a sun-like star gave equatorward migration \citep[see Fig.\,5 in][]{Kea00}.}

\begin{figure}
\begin{center}
\includegraphics[width=0.45\textwidth]{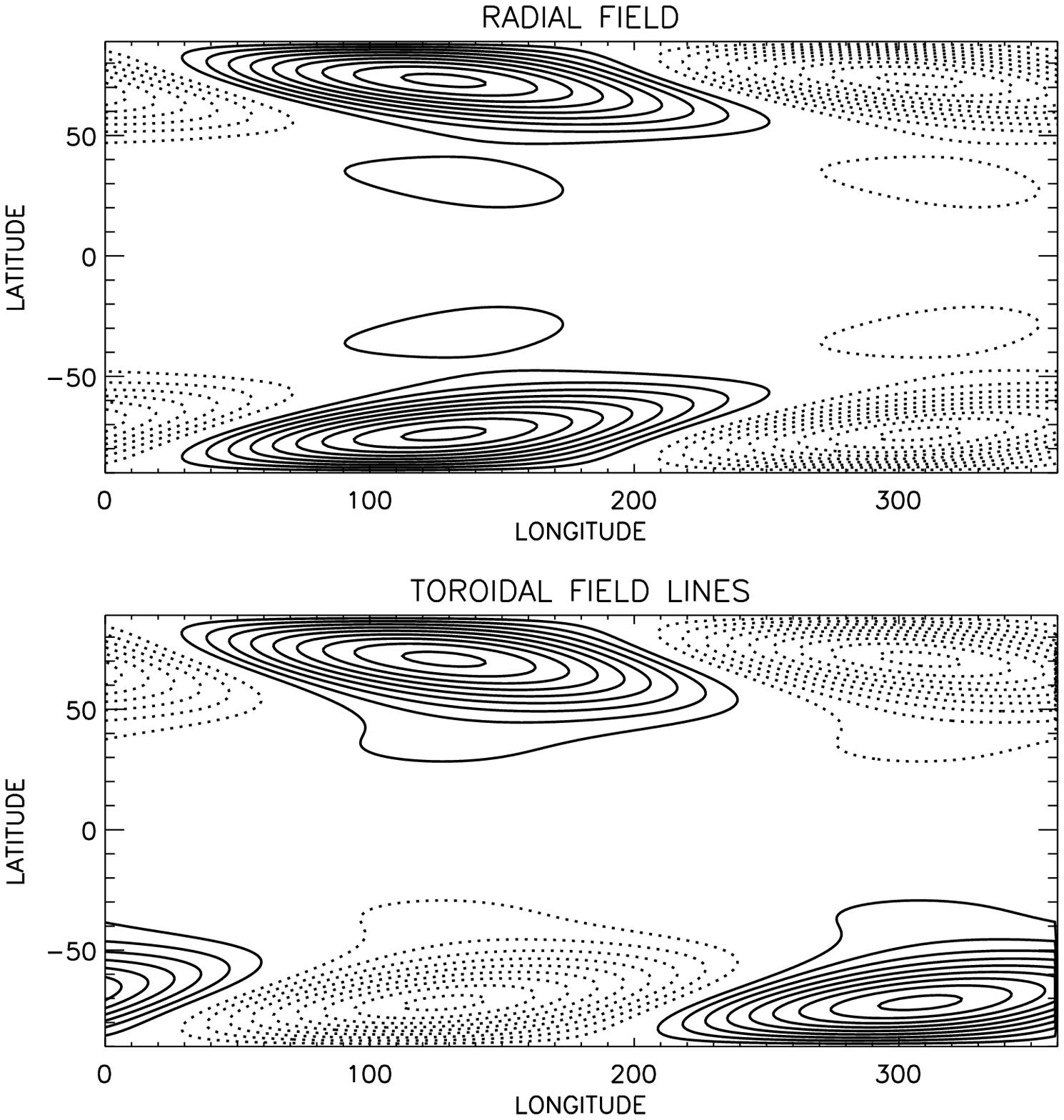}\\
\includegraphics[width=0.45\textwidth]{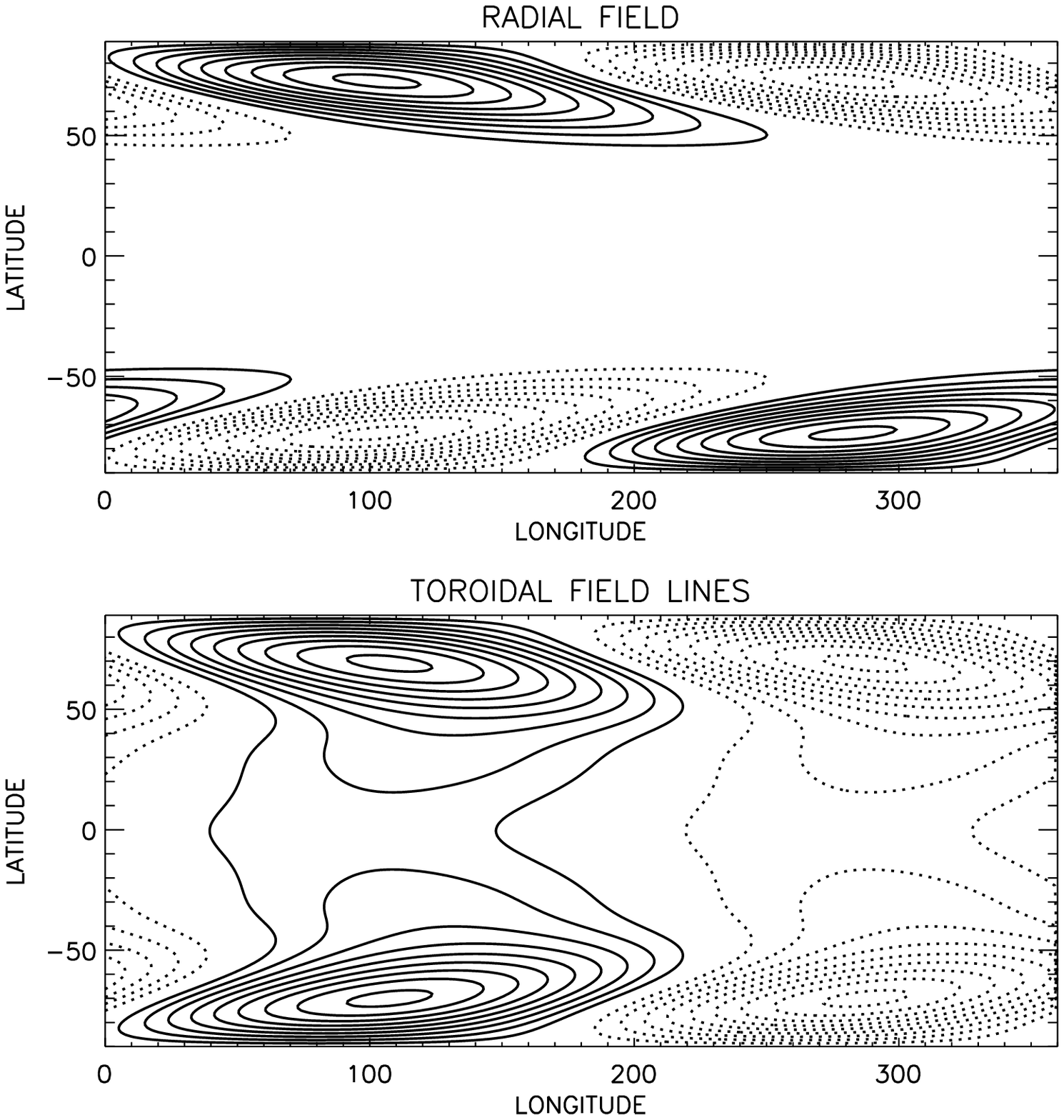}
\end{center}
\caption{\label{latlon}
Configuration of nonaxisymmetric modes in latitude-longitude coordinates: upper pair -- mode A1, lower pair - mode
S1. Pm = 1.3.}
\end{figure}

As butterfly diagrams are irrelevant to nonaxisymmetric modes, we represent
their field configurations by a diagram in
latitude-longitude coordinates (Fig.~\ref{latlon})

Of course, such figures have previously been presented in the literature,
but perhaps with not so much detailed inter-comparison.  Our aim is
to discuss to what extent the four dynamo modes under discussion can be
distinguished observationally at present time or in the
visible future and how it  could in principle be done.

\section{Distinguishing dynamo modes observationally}
\label{observ}

First of all, no single mode from the above list has a configuration similar to the solar case,
i.e. an oscillating axisymmetric mode of dipolar symmetry with equatorward
propagation, and with toroidal magnetic field (i.e. stellar spots) located near to the stellar equator.

The following point looks instructive. The toroidal magnetic field
becomes strong near to the stellar equator only for the
mode S0. In other
words, confirmed observation of stellar spots near the stellar equator is
a diagnostic feature for the mode S0.
Note however that, as can be seen from Fig.~3,
more or less strong equatorial spots can also be observed  for
A0 configurations from some inclinations.
The exact latitudinal pattern of toroidal field is likely to vary with the
angular form of $\alpha$, e.g. $\alpha \propto \sin^2\theta\cos\theta$ would push
it towards the equator
(simply because the maximum of one of the dynamo drivers becomes closer to the stellar equator).
In density-stratified convection, it seems likely
that helicity (which is often used as a proxy of the $\alpha$-effect) is indeed stronger near to the equator \citep{gast12}.
Excitation of mixed parity solutions is known for nonlinear spherical dynamos
\citep[e.g.][]{Betal89,JW91}
and for solar activity at the end of
the Maunder minimum \citep{snr94}, and
would also probably complicate interpretation.

Modes S1 and A1 have a pronounced longitudinal modulation of their nonaxisymmetric toroidal fields.
We note that photometric data for the Ca lines obtained in the framework of the
H-K project allow recognition of  rotation
modulation for some stars  \citep{betal95}.
Remarkably \citep{Ketal10}, this modulation is well-pronounced not for the
stars where the activity
cycle is strong (``excellent'' in the classification of \citet{betal95}),
but rather for the stars where cycles are moderate (``good'' of Baliunas et al.'s
 classification). If such modulation were to be isolated for
flare and flat stars in the classification of \citet{betal95},
i.e. for stars without marked cycles,  these
stars could be considered to have magnetic fields of
S1 or A1 configuration.

In order to distinguish between S and A modes it is necessary to compare
the magnetic field direction in ``Northern'' and ``Southern'' hemispheres.
The problem is that to perform this test observationally a sample of M dwarfs with rotation axis
having large inclination angle with respect to the line-of-sight
(ideally perpendicular) must be chosen and compared.
Recent development of Zeeman-Doppler Imaging (ZDI) has provided us with magnetic maps of a number of stars.
These maps can already be used to test at least some dynamo scenarios.
For instance, \citet{2008MNRAS.384...77M}, \citet{2008MNRAS.390..567M}, \citet{2008MNRAS.390..545D}, and
\citet{Mea10}
explored the surface magnetic field geometry and its evolution in a sample of M dwarfs consisting of partly and fully
convective objects. Only circularly polarized Stokes V spectra were used in this analysis. Therefore, the magnetic
maps presented in these studies reflect mostly the large-scale
magnetic structures that do not cancel out in
polarized light (if a sufficient number of observed time-series is available to
cover the full rotation cycle of the star uniformly).

The main results of the ZDI studies are the existence of a switch in the magnetic field geometry between
partly and fully convective M dwarfs: partly convective stars tend to host non-axisymmetric fields with
dominant toroidal components, whereas axisymmetric dipole-like fields are found in
fully convective stars.
The distribution of field topologies, however, was found to be even more complicated.
In particular, objects with weak toroidal fields were found among rapidly rotating fully convective stars.
This pointed towards a possible dipolar breakdown so the stars with similar stellar parameters
can host either dominant poloidal or toroidal configurations (see Fig.~15 in \citet{Mea10}).
A possible explanation for these observational findings was discussed in \citet{2013A&A...549L...5G}
who interpreted the magnetism of late M dwarfs as a result of a dynamo bistability and
the choice of the final dynamo state in the models discussed above depends on initial
conditions and does not change with time. We note here that, because our dynamo models are linear, 
they cannot provide evidence for bistability: the choice of the final dynamo state is fixed by
the initial conditions.
It is important to stress that no signature of a global magnetic topology change in individual stars has been seen.
It may be that much longer monitoring is required to observe such changes (which, in fact, would be a crucial test 
for the bistability scenario).
Note however that the activity
cycles in M dwarfs are expected to be much longer than that of the Sun.
\citet{ketal14} suggested a model for the observed
change of state that depends on the underlying activity cycle.

We note here some further points concerning the bistability problem. \cite{2013A&A...549L...5G}
(see also \cite{schrin12}) reported dynamo bistability at low
Rossby $Ro$ number from a 3D model, i.e. the dynamo action either produces a strong stable dipolar
field without any significant time variation ($\alpha^2$-dynamo), or a much weaker field which may
possibly oscillate and be described by an $\alpha^2 \Omega$-dynamo. The choice of
the final stable attractor depends on the initial conditions, or the prior
physical conditions in the stars in their early stages after their formation. For
slower rotators ($Ro > 0.1$), a dipolar breakdown is
reported as strong dipole field become less likely when rotational influence on
the convective flow becomes secondary. In principle, our illustrative model includes the possibility of a steady
axisymmetric dynamo (dotted line in Fig.~1) and we recognize that further investigation may result in
identification of this possibility with the actual field configurations in M dwarfs.

We can now try to compare our model predictions with the available ZDI maps.
We first note that almost all stars with dominant non-axisymmetric toroidal components
from the final sample of \citet{Mea10}
demonstrate similarities with the dipole-like, non-axisymmetric A1 mode (see Fig.~\ref{latlon}).
However, the observed orientation of the meridional component of the magnetic field
does not agree with that predicted.
For example, the radial field component of EV~Lac
is consistent with that
of the A1 mode \citep[see][]{2008MNRAS.390..567M},
but the $B_{\phi}$ component does not seem to show the expected surface distribution.
If we ignore the $B_{\phi}$ component and concentrate only on $B_{\rm r}$,
then the signatures of mode A1 (which changes in sign along longitude)
are found among both the groups of partly and fully convective stars, such as, say,
GJ~1245~B, GJ~1156, DT~Vir, GJ~182 and some others.
Note that the A1 mode is only a plausible
explanation of the observed picture of the magnetic field distribution in the framework of the dynamo
modes that we consider in our study. Of course, this in no way proves that the actual magnetic field pattern
in these stars can be described by such a simple model.
Also note that our dynamo model may be not fully adequate
especially at the stellar surface and that we show toroidal magnetic field distributions slightly below it.
(Remember that our boundary condition is of zero
toroidal field at the surface -- see the short discussion in Sect.~2.)

On the other hand, the mode S1 never seems to be observed. Many objects demonstrate a distribution of radial field,
which is characterized by opposite polarities in Northern and Southern
hemispheres at all longitudes, while a change in polarity would be expected.
Thus this disagrees with predictions for both A1 and S1 modes. Two stars,
YZ~CMi and DX~Cnc, seem to show a possible transition between A1 and S1 modes during
subsequent observing epochs, but this results is not well constrained by the ZDI maps,
and also the distribution of $B_{\phi}$ does not agree with that
of  any of the predicted modes.

The detection of A0 and S0 modes is more difficult because
the existing ZDI maps better represent the large-scale magnetic structures and not
the small-scale structures that might be associated with stellar spots.
However, we note that objects with detected non-axisymmetric toroidal configurations
often show localized magnetic structures that could be characterized
as spots. Stars such as DS~Leo, DT~Vir, GJ~182, GJ~1156, and GJ~1245 show spotty patterns of
the distribution of $B_{\rm r}$ on their surfaces. These spotty structures evolve on time scales of
years, and thus can be used to construct butterfly diagrams.
Unfortunately, similarly to the cases discussed previously,
the observed distribution of $B_{\phi}$ does not agree with those predicted.
In some stars we see the appearance of
single large scale spots and not a group of spots with different polarities distributed along a given latitude,
while in others (e.g. DT~Vir, DS~Leo, and OT~Ser) very complex surface structures
are detected and these stars are certainly good candidates for future monitoring and the
detection of A0/S0 modes. In this context it looks attractive to suppose that the spots at M dwarfs are caused by
radial magnetic field rather by eruptions of subsurface toroidal magnetic field, as is the case on the Sun.

Another distinct feature of the observed magnetic field distributions is the presence of polar spots in the majority
of fully convective objects. These spots can have positive (DS~Leo, EQ~Peg~A) or negative (CE~Boo, AD~Leo, YZ~CMi, WX~UMa, GJ~51)
signs of $B_{\rm r}$. It is unlikely that we observe a group of spots because only a single polarity
is detected. Therefore, these magnetic polar spots look simply like poles of the large-scale dipolar field.
Note that these polar spots also change their shapes, as detected for e.g., YZ~CMi. This possibly implies
that also other stars with polar spots should show the same evolution and their non-detection is connected
with too short monitoring time.

It is interesting to note that the formation of polar spots has recently been modelled by \citet{2015A&A...573A..68Y}
who found that, under certain conditions, their distributed dynamo models can spontaneously generate large-scale
dark spots at high stellar latitudes. Thus, these models represent an alternative explanation of the
observed magnetic field geometries in rapidly rotating low mass stars.

We may conclude at this point that the signatures of A0 and S0 modes, i.e. the presence of equatorial spots,
can indeed be found in M dwarfs with strong toroidal fields, but it is hard to see stable patterns
in their evolution. This is plausibly because the observations of these stars
do not extend over long enough time intervals.

It follows from the above considerations that it is not possible to
distinguish unambiguously
between different dynamo modes from the available ZDI maps. The main limitation is the
absence of long-time monitoring (i.e over several years).
On the other hand, it is
already possible to see a clear distinction between, say, A1 and S1 modes in some stars,
at least from the analysis of the radial field component. Additional
monitoring is strongly needed.

In order to put tighter constraints on the possible dynamo modes, observations of
stars with large inclination angles ($60^\circ-70^\circ$) are needed.
Among stars from \citet{Mea10}, there are many
that already meet this requirement, but none of them has
an inclination larger than $70^{\circ}$. Nevertheless, the construction
of the sample of stars with sufficiently large inclination angles does not
seem to be problematic. The only difficulty is that inclination angles are often
derived from ZDI itself so it is impossible to know these angles before
actual phase-resolved observations. If rotational periods are known (say, from analysis
of observed light curves) then inclinations can be derived from spectroscopically
known $\vsini$ values. In many cases, however, spectroscopic $\vsini$ values are
not accurately constrained which thus results in large errors in inclination.
Interferometry is an alternative technique
that may provide model independent estimates of inclination angles, but its
abilities (in most cases) are not yet sufficient to resolve the surfaces of small and faint M stars.

Despite the detected changes in the magnetic field topology from Stokes V spectra,
no strong evidence of the same changes have been noticed from the analysis of unpolarized
light, as discussed in \citet{2014A&A...563A..35S}.
The authors explored the distribution of magnetic filling factors on surfaces of four M dwarf stars
(from the very magnetically sensitive lines of the FeH molecule) and found similar distributions in all of them.
Note that no information about the magnetic field orientation could be derived from this study.
The only trend the authors could detect was in the distribution of filling factors and the strength of the
surface magnetic field, which seem to depend on the rotation rate. However this is still inconclusive because of
the small sample of stars used.

The only way to distinguish unambiguously between dynamo modes is by
monitoring
individual M dwarfs in polarized light by all possible means,
in order to construct time-series of ZDI maps. By tracking the evolution
of magnetic (and possible temperature) structures the corresponding butterfly diagrams could be constrained and
comparison made directly  with model predictions.
Fig.~\ref{dwarfdyn} tells us that the monitoring should be quite long,
over about 30 years (to be compared with
the 11-year solar cycle);  15-year monitoring would be sufficient for preliminary conclusions.

Normally, ZDI techniques allow constraint of a few spherical
harmonics of the surface magnetic field, with a typical maximum spherical
harmonic degree of about $5$-$10$. Thus, if the dipole has a
typical oscillatory cycle of $60$ years then it is very plausible that higher
degree modes can vary on shorter timescales. For instance, in geodynamo
($\alpha^2$) models, the timescales of secular variations of the magnetic field obey a $1/\ell$
law \citep{2012GeoJI.190..243C}. High-quality observations and time-tracking over $5$-$10$ years would then be
sufficient to see a notable secular variation of the magnetic field in the higher
spherical harmonic degrees. It looks thus promising in this respect to observe
rapidly rotating stars such as, say, V374~Peg \citep[P$=0.44$d, ][]{detal06,2008MNRAS.384...77M}.
Although no change in the magnetic field topology was reported between the two years of
observations (2005 and 2006), continuing monitoring of this star is strongly recommended because
the possible cycle is expected to be short.

On the other hand, a significant amount  of information concerning magnetic configurations can be obtained using DI
only, cf. that the solar magnetic cycle was isolated by Schwabe in the XIXth century before Zeeman
splitting in sunspots was observed.
DI techniques have existed for about 30 years but the time resolution of the published maps of M dwarfs at best only span
several years. Note that the length of activity cycles is expected to scale with
rotation rate. This issue was seems to have been first discussed by
\citet{netal84}. The tendency is to some
extent confirmed by observations in the Wilson program \citep{sb99}.
Observing the most rapidly rotating
stars may be a relatively easy task. For instance,
the rapidly rotating YZ~CMi, DX~Cnc, GJ~1156, GJ~51, etc.
are potentially good targets with which to begin.
In addition, the strength of the surface magnetic field clearly correlates with the rotation rate:
it grows steadily with decrease of rotation period until a saturation limit
is reached \citep{2009ApJ...692..538R}. Importantly, as was shown in \citet{2003A&A...397..147P} and more
recently in \citet{2014arXiv1408.6175R}, the magnetic field strength in the
non-saturated regime is a function only of rotation
rate and does not depend on any other stellar parameters (such as mass or radius), while the saturation limit
depends on the bolometric luminosity and thus differs from star to star. Therefore, studying rapid rotators
could produce interesting constraints on dynamo action in M stars.

So far all ZDI maps have been obtained by using observations in Stokes V only.
In order to resolve more surface detail, mapping of both linear
and circular polarization is needed. This imposes very strong observational
constraints because, e.g., long integration
times would be needed to collect sufficient signal in linearly polarized light.
High spectral resolution is also an essential requirement.
The instruments available for M dwarf research
are ESPaDOnS@CFHT ($3.6$~m, $R=65\,000$, $370$--$1005$~nm), NARVAL@TBL ($2$~m, $R=65\,000$, $370$--$1005$~nm)
\citep{1997MNRAS.291..658D},
HARPSpol@ESO ($3.6$~m, $R\approx110\,000$, $378$--$691$~nm) \citep{2011ASPC..437..237S,2011Msngr.143....7P}, and
HiVIS@AEOS ($3.7$~m, $R\approx50\,000$, $500$--$1000$~nm; $R\approx33\,000$, $1000$--$2500$~nm)
\citep{2003SPIE.4841.1115T,2006PASP..118..845H,2008PASP..120...89H},
as well as future instruments such as PEPSI@LBT ($2 \times 8.4$~m, $R=120\,000$, $383$--$907$~nm)
\citep{2011AN....332..753I}, SPIRou@CFHT ($3.6$~m, $R\approx70\,000$, $980$--$2350$~nm) \citep{2011ASPC..448..771A},
and CRIRES@VLT ($8$~m, $R=100\,000$, $1000$--$5300$~nm) \citep{2004SPIE.5492.1218K}.
Observing at infrared wavelengths with
high spectral resolution is superior to visual observations because of
the  cool temperatures of M dwarfs, and that the scaling of Zeeman
splitting is proportional to $\lambda^2$.
Thus instruments operating in the near-infrared would contribute greatly to studying
the evolution of magnetic topologies in M dwarfs.

\section{Discussion and conclusions}

Based on a relatively simple and standard mean field dynamo model for M dwarfs
we have presented four dynamo generated magnetic
configurations which can be excited in the framework of the model.
In this model, the parameter which determines the excited dynamo mode is 
the magnetic Prandtl number $P_{\rm m}$. Note that we identify 
the excited mode with the mode with the lowest marginal dynamo number. 
We appreciate that for more supercritical non-linear models 
this simple idea may become inadequate. 
However, we intend to begin our analysis with simple models and improve them
if observations will so require.

We started from the conventional expectation that observations and
dynamo theory are still quite remote from each other,
and comparison is a very uncertain and problematic undertaking.
After making the comparison it appears plausible
that the situation may not be quite as difficult as expected and
some preliminary identifications may be obtained immediately. We see
that at least some of the configurations look closer to observations
than others. In particular,  signatures of the S1 configuration
appear never to have been observed, while prospects for identification of the
A1 configuration perhaps look more promising.
The surface distribution of radial
magnetic field looks rather similar to the predictions of dynamo theory
while the distribution of toroidal components in
dynamo models appears quite different to the observational data.

Note that systematic investigation of magnetic activity of M dwarfs can be useful for the understanding of stellar
hydrodynamics.
Indeed, Eq.~(\ref{C_alpha}) suggests that dynamo action becomes stronger for stars with larger stellar radius,
if other governing parameters are unchanged.
In practice however turbulent diffusivity increases with stellar radius so the stellar hydrodynamical model predicts
that dynamo action becomes weaker for more massive stars. On the other hand,
Fig.~\ref{diff} tells us that preferred magnetic fields configurations are specific for weaker and stronger dynamo action, as
given by the $C_\alpha$ parameter. Confronting these expectations with future observations we may hope to deduce
the actual scaling of turbulent diffusivity with stellar radius.

In principle, we could start fitting dynamo models in order to reproduce an
improved phenomenology for $B_\phi$.
However our feeling is that, taken overall, the results on the topic
are still not stable enough, and it would be better to
discuss the available and forthcoming results in the framework of
the strategy suggested here,
as well as bearing in mind the
possibilities discussed above when analyzing any forthcoming
and improved dynamo modelling.
In any case, it looks plausible that the joint effort of observers and
dynamo modellers will be able to clarify the problem in the foreseeable future.

We recall the brief overview of dynamo modelling given
in Sect.~\ref{intro}, and appreciate that
the dynamo model we have used can of course be criticized on a number of grounds.
However it seems remarkable that all dynamo driven
magnetic field configurations for M dwarfs obtained in its framework
give a magnetic field that is concentrated at quite high
latitudes, i.e. they predict magnetic stellar activity in
regions that are far from the stellar equator.
Based on solar experiences, this result is quite
unexpected and deviates substantially from expectations.
On the other hand, the formation of high-latitude spots in rapidly rotating
fully-convective stars is also predicted by self-consistent global dynamo models
presented by \citet{2015A&A...573A..68Y}. This may indicate that the dynamo mechanism in
cool rapidly rotating stars is quite different from the solar case.

We finally stress again that our intention has been to explore
a possible methodology for investigation of the phenomenology of M dwarf
magnetism. We have illustrated this by reference to a particular dynamo model,
but as  more sophisticated and definitive models for the magnetic fields
are developed, the principles will remain relevant.

\section*{Acknowledgments}
D.~Shulyak acknowledges financial support from SFB~963~--~Astrophysical Flow Instabilities and Turbulence
(projects A16 and A17). D.~Sokoloff acknowledges partial support from RFBR under grant 12-02-00170-a. LK acknowledges support under RFBR grant 13-02-00277.
The authors are grateful to an anonymous referee for a detailed and constructive report which contributed to improvement of the paper.

\label{lastpage}

\end{document}